\documentclass[]{jingdong}

\definecolor{jsonstr}{rgb}{0.5,0.5,0.5}
\usepackage{authblk} 

\usepackage{marvosym}
\usepackage{listings}
\usepackage{xcolor}
\usepackage{booktabs}
\usepackage{colortbl}  
\usepackage{xcolor}    
\lstdefinelanguage{json}{
    basicstyle=\ttfamily\tiny,
    showstringspaces=false,
    breaklines=true,                 
    breakatwhitespace=false,
    frame=single,                    
    rulecolor=\color{gray!40},
    backgroundcolor=\color{gray!5},  
    literate=
     *{:}{{{\color{black}{:}}}}{1}
      {,}{{{\color{black}{,}}}}{1}
      {\{}{{{\color{black}{\{}}}}{1}
      {\}}{{{\color{black}{\}}}}}{1}
      {[}{{{\color{black}{[}}}}{1}
      {]}{{{\color{black}{]}}}}{1},
    morestring=[b]",
    stringstyle=\color{jsonstr},
}

\usepackage[utf8]{inputenc}
\usepackage[T1]{fontenc}

\usepackage{amsfonts}
\usepackage{amsmath}
\usepackage{amssymb}
\usepackage{algorithm}
\usepackage{algpseudocode}
\usepackage{booktabs}
\usepackage{xcolor}
\usepackage{enumitem}
\usepackage{inconsolata}
\usepackage{listings}
\usepackage{nicefrac}
\usepackage{siunitx}
\usepackage{url}
\usepackage{xspace}
\usepackage{wrapfig}
\usepackage{graphicx}

\crefformat{section}{\S#2#1#3}
\Crefformat{section}{\S#2#1#3}
\crefmultiformat{section}{\S#2#1#3}{ and \S#2#1#3}{, \S#2#1#3}{ and \S#2#1#3}
\Crefmultiformat{section}{\S#2#1#3}{ and \S#2#1#3}{, \S#2#1#3}{ and \S#2#1#3}
\crefrangeformat{section}{\S#3#1#4 to \S#5#2#6}
\Crefrangeformat{section}{\S#3#1#4 to \S#5#2#6}

\newcommand{\newpara}[1]{\vspace{0.1cm}\noindent\textbf{#1}}

\definecolor{templatekeyword}{HTML}{E1251B}
\definecolor{templatestring}{HTML}{0B7A75}
\definecolor{templatecomment}{HTML}{6B7280}
\definecolor{templatecodebg}{HTML}{F7F7F8}

\lstdefinestyle{templatecode}{
  basicstyle=\ttfamily\small,
  columns=fullflexible,
  backgroundcolor=\color{templatecodebg},
  frame=single,
  rulecolor=\color{black!15},
  numberstyle=\tiny\color{gray},
  keywordstyle=\color{templatekeyword},
  commentstyle=\color{templatecomment},
  stringstyle=\color{templatestring},
  showstringspaces=false,
  tabsize=2,
  breaklines=true,
  breakatwhitespace=true,
  captionpos=b,
  xleftmargin=3.4pt,
  xrightmargin=3.4pt
}

\title{JoyAI-Voice 2.0: A Full-Continuous Autoregressive Speech Generation Model with Semantic-Acoustic Joint Representation}

\vspace{-40pt}

\affiliation{Joy Future Academy, JD.com}

\vspace{-40pt}

\abstract{
We present JoyAI-Voice~2.0, an end-to-end anthropomorphic speech generation model built upon a fully continuous, dual-encoder architecture. Raw speech is encoded into continuous latents and partitioned into patches. Each patch is decomposed by a semantic-acoustic dual encoder into a semantically purified representation and an acoustic representation, which are fused and jointly fed to a causal autoregressive Transformer for planning. The Transformer predicts the conditioning for the next patch, and a local diffusion Transformer renders its full latents for 48\,kHz synthesis. The model is trained with a joint flow-matching and stop-prediction objective, followed by supervised fine-tuning and reinforcement learning with DiffusionNFT to improve model performance. It achieves the lowest average word error rate of 2.51\% on Seed-TTS, a 14.9\% relative reduction over the strongest baseline, and state-of-the-art attribute fidelity on InstructTTSEval in both Chinese and English, leading on 5 of 10 perceptual dimensions with the highest overall score of 0.893 on MDVD-Eval.

\vspace{-30pt}
}

\begin{document}

\maketitle


\begin{figure}[H]
\centering
\begin{subfigure}[t]{0.49\textwidth}
    \centering
    \includegraphics[width=\linewidth]{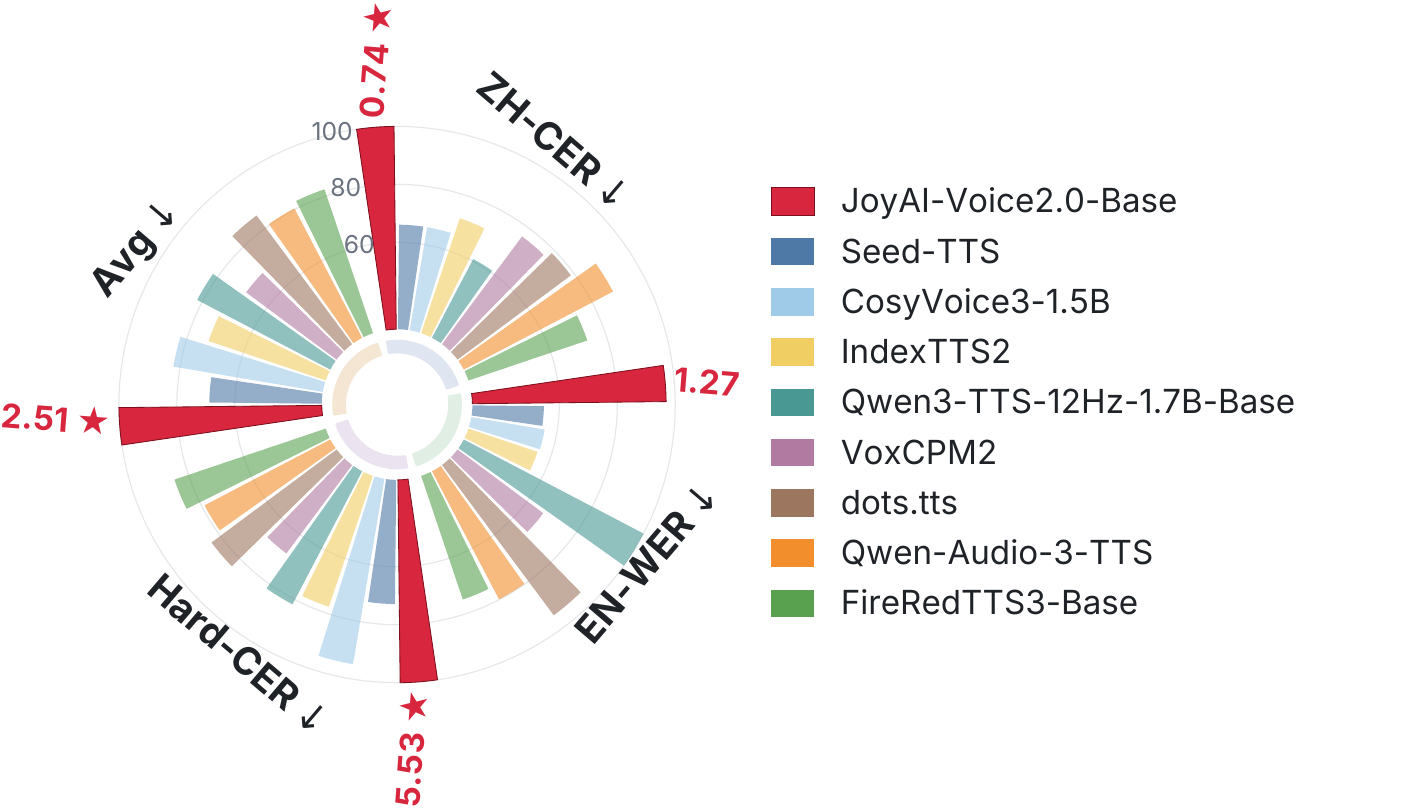}
    \caption{Voice cloning}
    \label{fig:result1}
\end{subfigure}
\hfill
\begin{subfigure}[t]{0.49\textwidth}
    \centering
    \includegraphics[width=\linewidth]{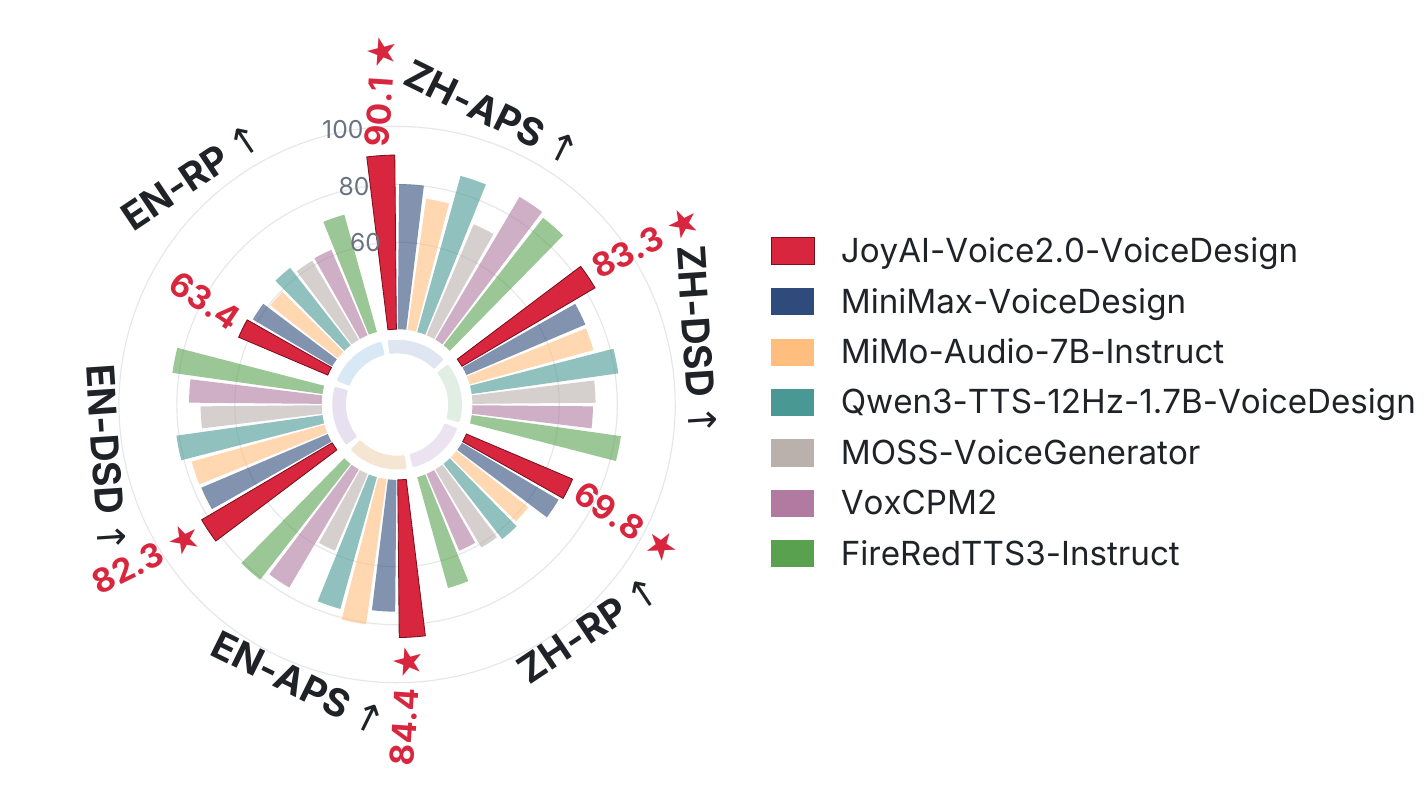}
    \caption{Voice design}
    \label{fig:result2}
\end{subfigure}
\caption{Benchmark results of JoyAI-Voice~2.0 on TTS evaluation: (a) voice cloning on Seed-TTS, achieving the lowest average WER of 2.51\%; (b) voice design on InstructTTSEval, attaining state-of-the-art attribute fidelity in both Chinese and English.}

\label{fig:results}
\end{figure}

\begin{figure}[H]
\centering
\includegraphics[width=0.6\textwidth]{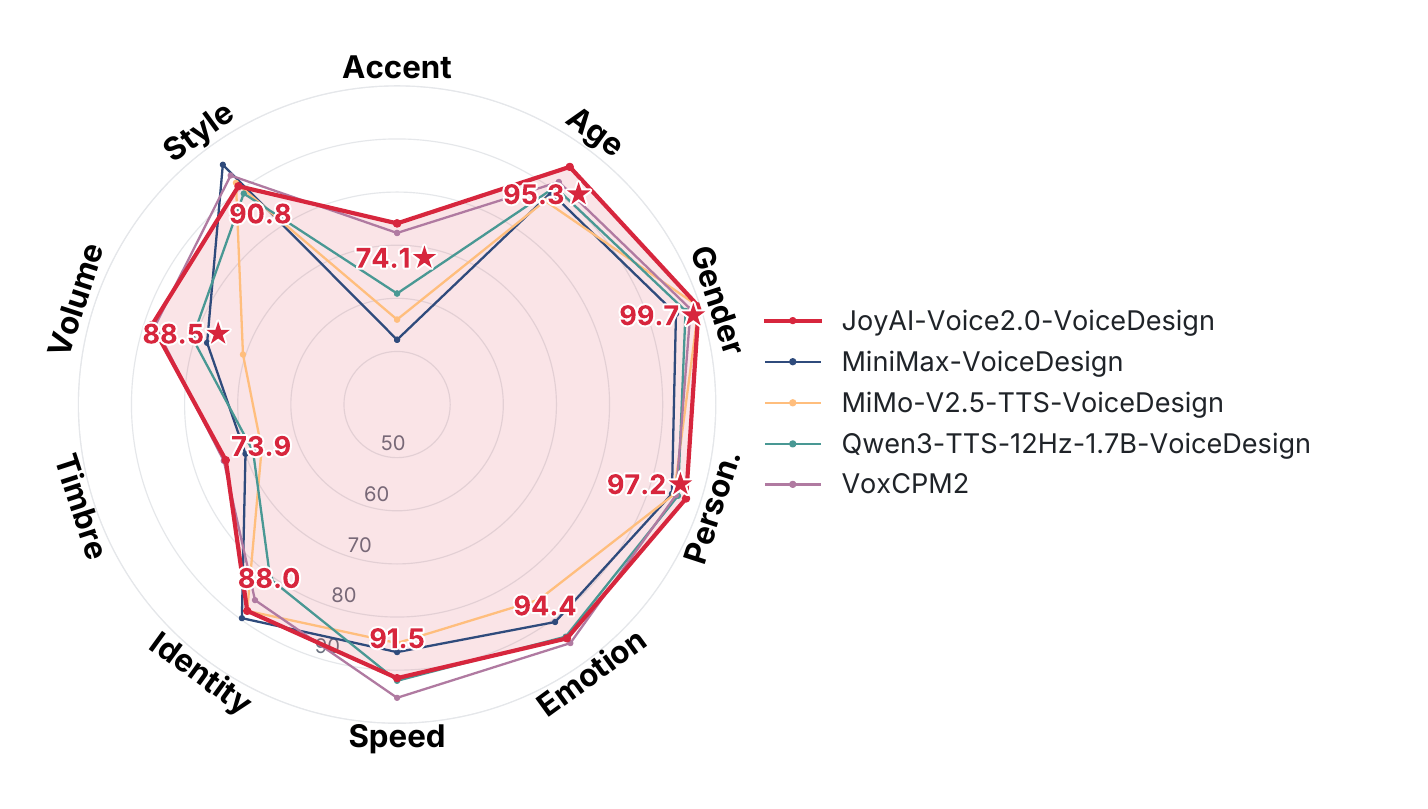}
\caption{Multi-dimensional instruction following performance on MDVD-Eval.}
\label{fig:result3}
\end{figure}
\section{Introduction}
\label{sec:introduction}

Text-to-Speech (TTS) aims to synthesize natural and intelligible speech from input text. Recent work based on large language models (LLMs) and diffusion models has substantially advanced zero-shot voice cloning and speech generation~\cite{valle2025,cosyvoice2024,seedtts2024}. Depending on the form of the generation target, existing systems follow three main paradigms: discrete-token autoregressive modeling~\cite{valle2025,cosyvoice2_2024,seedtts2024}, non-autoregressive diffusion generation~\cite{voicebox2023,ns3_2024,f5tts2025}, and continuous-representation autoregressive modeling~\cite{ditar2025,voxcpm2025,dotstts2026}.

Among these paradigms, we adopt \textbf{continuous-representation autoregressive modeling}. Discrete-token systems rely on audio tokenizer that quantize waveforms into codebook indices~\cite{soundstream2021,encodec2022}. This quantization bottleneck irreversibly discards timbre micro-textures and paralinguistic details, and reconstruction studies have shown that discrete codecs systematically trail continuous representations in perceptual audio quality~\cite{dotstts2026}. In our previous work JoyVoice~\cite{joyvoice2025}, we built an end-to-end TTS model upon discrete speech tokens and observed this ceiling directly: the expressiveness of the synthesized speech is bounded by the codec's representation capacity. A growing body of work, including DiTAR~\cite{ditar2025}, VoxCPM~\cite{voxcpm2025}, and dots.tts~\cite{dotstts2026}, demonstrates that continuous speech representation can break through this limit by removing quantization altogether. We therefore upgrade the discrete representation of JoyVoice~1.0 to a fully continuous one.

We further select the autoregressive rather than non-autoregressive structure. NAR diffusion systems excel at single-pass offline synthesis~\cite{voicebox2023,ns3_2024}, but autoregressive models natively support incremental, prompt-conditioned streaming synthesis that consumes upstream text streams as they are emitted---a core requirement for real-time conversational deployment. We therefore target an autoregressive design with streaming emission and millisecond-level first-packet latency.

However, continuous-representation autoregressive models carry each step's deviation into the next-step conditioning, and error accumulates over long rollouts, making training and generation highly unstable~\cite{dotstts2026}. To address this, we introduce a semantic-acoustic joint representation, which decomposes each patch of continuous latents into two complementary continuous representations before planning. The semantic representation is distilled from a pre-trained ASR model and remains frozen throughout training, preserving a stable, noise-purified semantic anchor that shields the autoregressive planner from acoustic nuisances and long-range error feedback. The acoustic representation is then learned to complement this anchor with the fine-grained details it discards, so that all information discarded for stability is recovered through a separate pathway rather than sacrificed. This divide-and-conquer design secures planning stability on the semantic side while raising the fidelity ceiling on the acoustic side, substantially improving both training stability and model performance.

Building on these designs, we present \textbf{JoyAI-Voice~2.0}, an end-to-end speech generation model built upon a fully continuous dual-encoder autoregressive architecture. The key capabilities are summarized as follows:

\begin{itemize}
    \item \textbf{Fully Continuous Autoregressive Modeling with a Dual-Encoder Architecture.} The model consumes text tokens and drives a per-patch flow-matching DiT head~\cite{dit2023,lipman2023} over continuous AudioVAE latents~\cite{voxcpm2025}, eliminating discrete speech tokenization entirely. A semantic-acoustic dual-encoder decomposes each patch into complementary continuous representations, which are fused and jointly fed to the language model for planning, yielding stability comparable to discrete systems without any quantization loss.
    \item \textbf{Instruction-Based Voice Design.} The model creates new timbres from natural-language descriptions and performs fine-grained control over emotion, speaking style, and accent within a single generation pass, without separate controllability modules or post-hoc editing. The voice design variant attains state-of-the-art attribute fidelity on InstructTTSEval in both Chinese and English, and leads on 5 of 10 perceptual dimensions with the highest overall score of 0.893 on our proposed MDVD-Eval.

    \item \textbf{Zero-Shot Voice Cloning.} Given only a few seconds of speech from the target speaker, the model preserves speaker identity, style, and paralinguistic characteristics while remaining robust to moderate reference noise. On the Seed-TTS benchmark, it achieves an average word error rate of 2.51\%, a 14.9\% relative reduction over the strongest baseline. Paralinguistic tags~\cite{jdtalker2026} such as \texttt{[Laughter]} and \texttt{[Sigh]} can be inserted at arbitrary positions to trigger non-verbal vocal events at specified moments.

\end{itemize}

The remainder of this report is organized as follows. Section~\ref{sec:architecture} details the continuous-space dual-encoder architecture, including the semantic-acoustic joint representation, the causal autoregressive Transformer, and the autoregressive diffusion Transformer. Section~\ref{sec:training} introduces the training methodology, covering pre-training, supervised fine-tuning, and reinforcement learning. Section~\ref{sec:experiments} reports comprehensive experimental results, including data and evaluation setup, and the performance of voice cloning and voice design. Section~\ref{sec:conclusion} finally concludes the report.

\section{Full-Continuous Dual-Encoder Architecture}
\label{sec:architecture}

\begin{figure}
\centering
\includegraphics[width=1.0\linewidth]{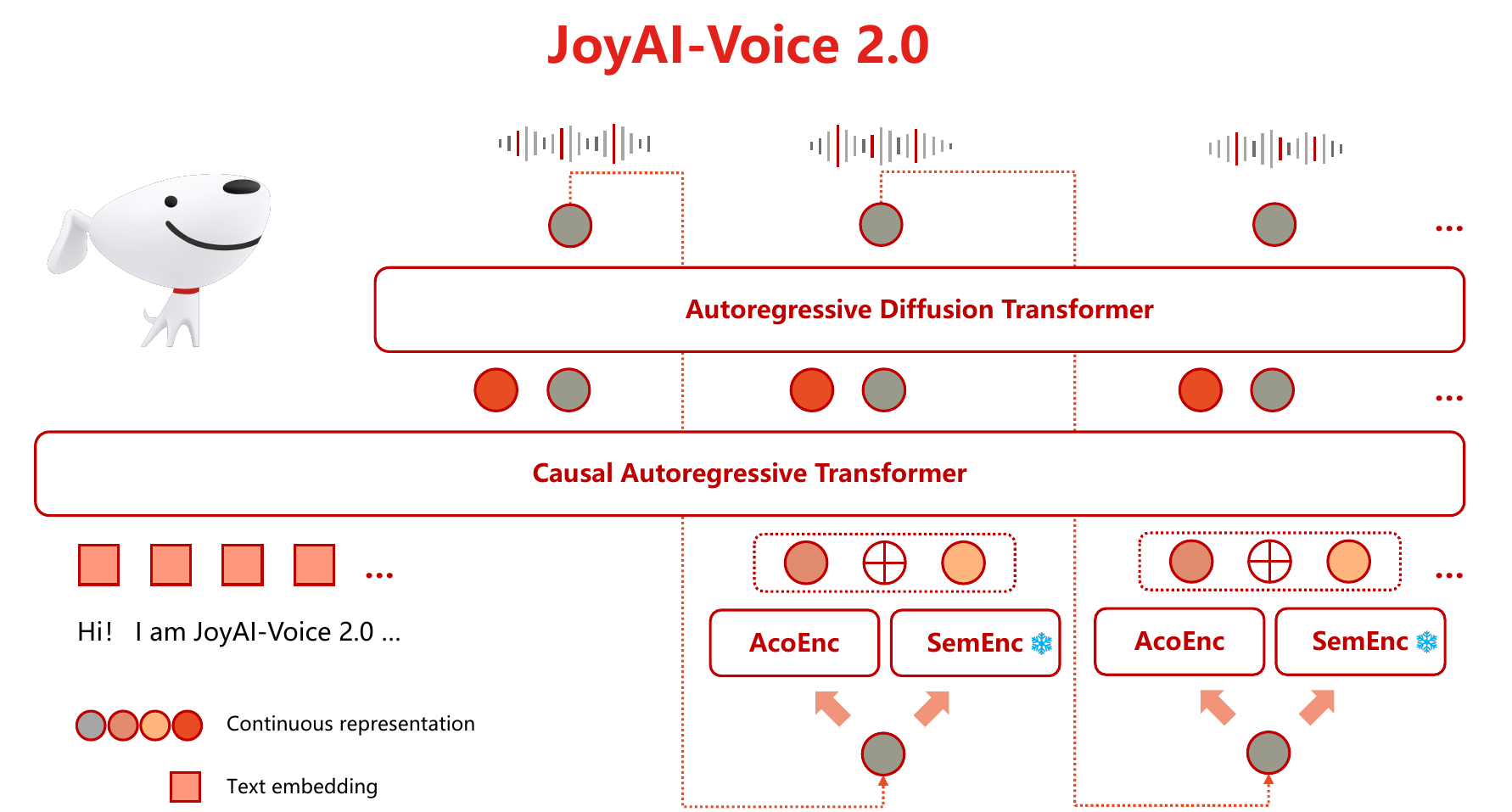}
\caption{Architecture of JoyAI-Voice 2.0. Raw waveforms are compressed into continuous latents by an AudioVAE and partitioned into patches. Each patch is decomposed by a dual encoder into a semantically purified representation and an acoustic representation, which are fused and jointly fed to a causal autoregressive Transformer for planning. A local diffusion Transformer renders the latents of the next patch, which are decoded into 48\,kHz waveforms.}
\label{fig:architecture}
\end{figure}

JoyAI-Voice~2.0 adopts a divide-and-conquer generation pipeline (Figure~\ref{fig:architecture}). Raw waveforms are compressed into continuous latents by an AudioVAE and partitioned into fixed-length patches. Each patch is processed by a semantic-acoustic dual encoder into a semantically purified representation and an acoustic representation, which are fused and jointly fed to a causal autoregressive Transformer for planning. The Transformer predicts the conditioning for the next patch, and a local diffusion Transformer renders its full latents, which are decoded into 48\,kHz waveforms. No discrete tokenization occurs anywhere in the pipeline, and all generation targets remain in the continuous latent space.

\subsection{Semantic-Acoustic Joint Representation}
\label{sec:dual-encoder}

The semantic-acoustic joint representation is derived from the continuous latents of an AudioVAE, which follows the architecture of VoxCPM2~\cite{voxcpm2_2026}. The encoder compresses raw waveforms into continuous latents at 25~Hz, and the decoder reconstructs 48\,kHz waveforms directly from these latents. The latent sequence is partitioned into non-overlapping patches of two consecutive frames, corresponding to a patch rate of 12.5~Hz.

Each patch is processed by two dedicated encoders into complementary continuous representations. The semantic representation is a transcription-level summary distilled by the semantic encoder, which filters acoustic noise and task-irrelevant details, improving the stability of autoregressive planning over long rollouts. The acoustic representation is an information-dense aggregation produced by the acoustic encoder, which preserves the spectral texture, speaker timbre, and micro-prosodic cues required for high-fidelity rendering. The two representations are fused and jointly consumed by the language model, so that planning is conditioned on purified semantics complemented by complete acoustic information. 

\subsubsection{Semantic Encoder}
\label{sec:semantic-encoder}

The semantic encoder (SemEnc) is built on the pre-trained ASR model Whisper-large-v3~\cite{whisper2023}, following the MM-Tokenizer design of JoyVoice~\cite{joyvoice2025}. A causal convolutional front-end projects the VAE latents into the backbone's hidden dimension and downsamples them by two, aligning the latent stream with the patch rate. Supervision is restricted to the ASR transcription objective, so the learned representation excludes acoustic nuisances. Within each patch, the hidden states of the two latent frames are pooled into the semantic representation. The ASR decoder head is discarded at inference.

\subsubsection{Acoustic Encoder}
\label{sec:acoustic-encoder}

The acoustic encoder (AcoEnc) is a lightweight Transformer aggregator trained jointly with the rest of the model. It attends over the latent frames within each patch and pools them into a single vector in the unified hidden space of the autoregressive Transformer, preserving the fine-grained details that the semantic representation discards. 

\subsection{Causal Autoregressive Transformer}
\label{sec:ar-transformer}

The planning stage is a decoder-only causal Transformer built on Qwen2.5-1.5B~\cite{qwen25_2024}. The input sequence consists of a BPE-tokenized text prefix followed by the speech patches, where each patch position carries the fused semantic and acoustic representations projected into the unified hidden space via modality-specific linear adapters. At each step, conditioned on the text prompt and all preceding patches, the Transformer predicts the conditioning for the next patch. The patch-level autoregressive formulation preserves long-range coherence, and the causal design supports incremental streaming generation with low first-packet latency.

\subsection{Local Diffusion Transformer}
\label{sec:diffusion-transformer}

The rendering stage is a local diffusion Transformer (LocDiT) that generates the full latents of the next patch from the planned conditioning. Denoising operates only within the target patch, with all previously generated latents injected as prefix context, converting intra-patch generation into contextual outpainting. The network is trained with the flow-matching objective~\cite{lipman2023}, with timestep and conditioning injected via adaptive layer normalization following DiT~\cite{dit2023}. At inference, each rendered patch is decoded by the frozen VAE decoder into 48\,kHz waveforms, and the conditioning for subsequent patches is derived from previous outputs. The planner and renderer thus focus on long-range semantic coherence and local spectral fidelity, respectively.

\section{Training}
\label{sec:training}
\subsection{Pre-Training \& Supervised Fine-Tuning}
We pre-train the model to generate continuous acoustic patches autoregressively from text.
Training utterances are represented as sequences of AudioVAE latent patches, with text provided as a conditioning prefix.
Under teacher forcing, the autoregressive backbone uses the fused semantic and acoustic representations of preceding ground-truth patches to condition generation of the next patch.
A CAM++ encoder extracts the speaker embedding from a random 10-second crop of the training utterance; utterances shorter than 10 seconds are used in full.
We initialize the autoregressive backbone from Qwen2.5-1.5B and randomly initialize the diffusion Transformer.
The pretrained AudioVAE, speaker encoder, and semantic encoder backbone remain frozen during speech generation training.


\newpara{Training objective.} 
We jointly supervise acoustic generation with a patch-level flow-matching loss~\cite{lipman2023} and learn speech termination with an auxiliary classifier.
We use precomputed AudioVAE latents when available and otherwise extract them with the frozen encoder from 16\,kHz waveforms.
The text forms a conditioning prefix, followed by the speech patch sequence.
We perturb the AudioVAE latents with Gaussian noise of standard deviation 0.1 and then normalize them using corpus-level statistics.
Let $z$ denote the resulting normalized latent sequence; the same augmented latents provide the conditioning history and the data endpoint of the flow-matching path.
For each utterance, we sample a timestep $t\sim\mathcal{U}(0,1)$ shared across its patches and Gaussian noise $\epsilon\sim\mathcal{N}(0,I)$, and construct
\begin{equation}
\label{eq:supervised_flow_path}
z_t=(1-t)\epsilon+t z, \qquad u_t=z-\epsilon.
\end{equation}
The diffusion Transformer $v_{\theta}(z_t,t,c)$ predicts the target velocity $u_t$, where $c$ comprises the autoregressive hidden states, preceding acoustic patches, and the speaker embedding.
Each noisy patch can attend to its own patch tokens and preceding history, but not to other noisy patches or future clean latents.
An auxiliary binary end-of-sequence (EOS) classifier predicts whether the next generated patch is the final patch of the utterance.
The supervised objective is
\begin{equation}
\label{eq:supervised_training_loss}
\mathcal{L}_{\mathrm{sup}}
=
\mathbb{E}_{z,t,\epsilon}
\left[
\operatorname{MSE}
\left(v_{\theta}(z_t,t,c),u_t\right)
\right]
+
\mathcal{L}_{\mathrm{eos}},
\end{equation}
where the flow-matching error is averaged over valid acoustic patches and $c$ denotes their conditioning context.
The EOS loss uses class-balanced weighting within each utterance.
The EOS classifier takes detached autoregressive hidden states as input, so $\mathcal{L}_{\mathrm{eos}}$ updates only the classifier without propagating gradients into the backbone.

\newpara{Condition Dropout.}
To support classifier-free guidance (CFG), we independently drop the autoregressive hidden-state conditioning and speaker conditioning at the utterance level.
The hidden states are masked before projection into the diffusion Transformer, while the speaker embedding is masked after projection.
The text prefix remains visible to the autoregressive backbone, and preceding acoustic latents remain available to the diffusion Transformer.

\newpara{Training configuration.}
We use AdamW with bfloat16 mixed precision.
For efficient training, we implement block-structured attention in the diffusion Transformer with FlexAttention~\cite{guessous2024flexattention} and causal self-attention in the autoregressive backbone with FlashAttention~\cite{dao2022flashattention}.
Batches are formed by length bucketing under audio-duration and text-token budgets.

The pretrained model is then adapted to downstream generation tasks through supervised fine-tuning. We perform supervised fine-tuning (SFT) to obtain separate base and instruction-conditioned variants from a shared pretrained checkpoint, using the same flow-matching and stop-prediction objectives.
Base fine-tuning uses paired speech and text, while instruction fine-tuning additionally uses natural-language annotations of speaker characteristics, speaking style, and emotion.
When instruction annotations are available, we sample from the supported descriptions and prepend the selected instruction to the text.
The speech remains the prediction target; the instruction is provided as conditioning rather than as a text-generation target.
The same generation modules remain trainable, while the pretrained representation and speaker-encoding backbones stay frozen.

\subsection{Reinforcement Learning} 
TTS models often struggle to faithfully synthesize challenging text, such as long, repetitive sentences and tongue twisters, whose limited coverage in large-scale pre-training data can lead to hallucinations.
Imperfect speaker consistency in the training data may also limit voice-cloning fidelity.
After SFT, we use reinforcement learning with DiffusionNFT~\cite{zheng2025diffusionnft} to improve synthesis accuracy on long and challenging text and speaker similarity in zero-shot voice cloning.

\newpara{Training data and rollouts.}
To broaden the text patterns encountered during RL, we introduce repetition at the word, phrase, and clause levels, inspired by the text augmentation strategy in FlowTTS-GRPO~\cite{wang2026flowttsgrpo}.
For each target text $x$ and reference utterance $y_{\mathrm{ref}}$, we generate multiple speech samples with different noise initializations.
The reference speech and its transcript form the prompt, and the online policy model generates speech for the target text as a continuation of the reference audio.

RL rollouts follow the continuation-based zero-shot voice-cloning setting.
Given a reference utterance $y_{\mathrm{ref}}$, its transcript $x_{\mathrm{ref}}$, and a target text $x$, we sample $G$ speech continuations using different noise initializations:
\begin{equation}
\label{eq:rl_rollout}
y_i \sim \pi_{\mathrm{roll}}^{\mathrm{CFG}}
\left(\,\cdot \mid x, x_{\mathrm{ref}}, y_{\mathrm{ref}}\right),
\qquad i = 1,\ldots,G
\end{equation}
where $\pi_{\mathrm{roll}}^{\mathrm{CFG}}$ denotes the CFG-guided rollout policy and $y_i$ is the generated continuation corresponding to the target text.
We use the acoustic latents from CFG-guided rollouts to construct NFT training targets for the fully conditioned branch, with both text and speaker embedding provided.
During optimization, we maintain only the conditional branch, using CFG-generated samples as targets in the NFT objective.

\newpara{Reward function.}
Each generated utterance $y$ is evaluated using two rewards: negative word error rate (WER), computed between its ASR transcription and the target text, and the cosine similarity between speaker embeddings extracted from the generated and reference speech.
The combined reward is
\begin{equation}
\label{eq:rl_multi_reward}
R = \lambda_{\mathrm{asr}} R_{\mathrm{asr}} + \lambda_{\mathrm{ss}} R_{\mathrm{ss}}.
\end{equation}
Here, $R_{\mathrm{asr}}$ and $R_{\mathrm{ss}}$ denote the normalized ASR and speaker-similarity rewards, respectively.
Each reward is normalized independently by subtracting its within-group mean and dividing by its global standard deviation, computed over all samples in the current batch.
The weights $\lambda_{\mathrm{asr}}$ and $\lambda_{\mathrm{ss}}$ balance the two normalized rewards.

We observe that the ASR reward fluctuates in the later stages of RL and becomes less informative about further progress.
Speaker similarity, by contrast, provides finer-grained feedback for continued refinement of the generated voice.
We therefore place greater emphasis on content accuracy during the first few hundred training steps ($\lambda_{\mathrm{asr}}:\lambda_{\mathrm{ss}}=5:1$), then reverse the ratio to $1:5$ as the optimization focus shifts toward speaker fidelity.


\section{Experiments}
\label{sec:experiments}
\subsection{Data} 

To prepare millions of hours of training data, we developed a DAG-based distributed pipeline for large-scale audio processing, in which each stage---format standardization, background-music separation, speaker diarization, DNSMOS-based quality assessment, and transcription---is implemented as an independent, fault-tolerant node.
Audio is partitioned into packages, each containing a batch of long-form recordings and dispatched via a message queue.
Upon completion, each package is automatically routed to its downstream nodes according to the DAG dependencies.
Once the workflow is configured, the pipeline runs unattended, converting raw audio into training-ready records.

Within this pipeline, each package is processed in three stages.
First, preprocessing standardizes the audio format, separates background music, performs speaker diarization, and computes DNSMOS scores for quality filtering.
Second, the resulting speech segments are transcribed in parallel by multiple ASR systems whose outputs are cross-validated via pairwise WER comparison.
The most consistent transcription is retained, which substantially suppresses recognition noise compared with any single system.
In parallel, a speaker embedding model and a frozen AudioVAE encode each segment into speaker representations and continuous latents, respectively, matching the representations required by our model (Section~\ref{sec:architecture}).
Finally, the selected transcripts, speaker embeddings, VAE latents, and segment metadata are aligned and merged into structured training records.

\subsection{Evaluation} 

We evaluate JoyAI-Voice 2.0 along two dimensions: zero-shot voice cloning and voice design.

\begin{itemize}
    \item \textbf{Zero-Shot Voice Cloning}: We adopt the Seed-TTS test set, including the test-zh, test-en, and test-hard subsets, to assess the model's ability to clone voices. Content consistency is measured by character error rate (CER) for Chinese and word error rate (WER) for English, and speaker similarity (SIM) is computed using a WavLM-based speaker verification model.
    \item \textbf{Voice Design}: We adopt the InstructTTSEval benchmark~\cite{instructttseval2025}, which comprises three instruction types of increasing abstraction: Acoustic-Parameter Specification (APS), Descriptive-Style Directive (DSD), and Role-Play (RP). For each sample, an LLM judge determines whether the synthesized speech is consistent with the instruction, and we report the resulting instruction-following accuracy (\%) for each type. Since Gemini-2.5-Pro-Preview-05-06 used in the original benchmark has been deprecated, we substitute it with Gemini-2.5-Pro.
\end{itemize}

To further evaluate whether a model can simultaneously realize multiple attributes specified in a single instruction, we introduce \textbf{MultiDim-VoiceDesign-Eval (MDVD-Eval)}, covering 10 perceptual dimensions—accent, age, gender, personality, emotion, speed, identity, timbre, volume, and style—with each instruction combining 2 to 4 attributes across 1,000 Chinese cases. An LLM judge decomposes each instruction into its constituent attributes and scores whether each is audibly reflected in the synthesized speech, yielding a per-case atomic score as the fraction of matched attributes. To validate this protocol, we compare three candidate judges against human annotations from nine professional audio researchers along two granularities: \textit{Atomic Agreement} (per-attribute level) and \textit{Judge Agreement} (overall case-level verdict). As shown in Table~\ref{tab:judge_agreement}, Gemini-2.5-Pro achieves the highest Atomic Agreement ($79.9\%$) and is therefore adopted as the judge.

\begin{table}[ht]
\centering
\small
\caption{Judge--human agreement on MDVD-Eval, averaged over nine annotators. Higher is better.}
\label{tab:judge_agreement}
\begin{tabular}{lcc}
\toprule
\textbf{Judge Model} & \textbf{Judge Agreement} ($\uparrow$) & \textbf{Atomic Agreement} ($\uparrow$) \\
\midrule
Gemini-2.5-Pro         & 0.561 & \textbf{0.799} \\
Gemini-3.1-Pro-Preview & 0.635 & 0.731 \\
Gemini-3-Flash-Preview & 0.516 & 0.765 \\
\bottomrule
\end{tabular}
\end{table}

\subsection{Performance} 

\subsubsection{Voice Cloning}

Table~\ref{tab:zeroshot_cloning} compares JoyAI-Voice2.0-Base with leading zero-shot TTS systems on the Seed-TTS test set. JoyAI-Voice2.0-Base achieves the lowest average recognition error of $2.51\%$ among all evaluated systems. Compared with dots.tts~\cite{dotstts2026}, the strongest baseline in average recognition error at $2.95\%$, JoyAI-Voice2.0-Base reduces the error by $14.9\%$ relative. On test-zh, JoyAI-Voice2.0-Base obtains the lowest CER of $0.74\%$, substantially ahead of MiniMax-Speech~\cite{minimax2025} at $0.83\%$. On test-hard, it achieves the lowest CER of $5.53\%$, outperforming CosyVoice3-1.5B~\cite{cosyvoice3_2025} at $5.83\%$, demonstrating stronger robustness on challenging utterances. On test-en, Qwen3-TTS-12Hz-1.7B-Base~\cite{qwen3tts2026} attains the lowest WER of $1.23\%$, narrowly ahead of JoyAI-Voice2.0-Base at $1.27\%$. For speaker similarity, dots.tts achieves the highest SIM on test-zh, test-hard, and on average, while FireRedTTS3-Base attains the highest SIM on test-en ($77.2$ vs.\ $77.1$ for dots.tts). JoyAI-Voice2.0-Base delivers competitive but not top-ranking SIM scores, suggesting room to further improve timbre and prosody fidelity.

\begin{table}[ht]
\centering
\caption{Zero-shot voice cloning performance on the Seed-TTS test set. CER, WER, and SIM ($\uparrow$/$\downarrow$ as noted) denote character error rate, word error rate, and speaker similarity, respectively. Best results are marked in bold.}
\label{tab:zeroshot_cloning}
\resizebox{\textwidth}{!}{
\begin{tabular}{lcccccccc}
\toprule
 & \multicolumn{2}{c}{\textbf{ZH}} & \multicolumn{2}{c}{\textbf{EN}} & \multicolumn{2}{c}{\textbf{Hard}} & \multicolumn{2}{c}{\textbf{Avg}} \\
\cmidrule(lr){2-3} \cmidrule(lr){4-5} \cmidrule(lr){6-7} \cmidrule(lr){8-9}
\textbf{Model} & CER ($\downarrow$) & SIM ($\uparrow$) & WER ($\downarrow$) & SIM ($\uparrow$) & CER ($\downarrow$) & SIM ($\uparrow$) & CER/WER ($\downarrow$) & SIM ($\uparrow$) \\
\midrule
Seed-TTS~\cite{seedtts2024}            & 1.12 & 79.6 & 2.25 & 76.2 & 7.59 & 77.6 & 3.65 & 77.8 \\
MaskGCT~\cite{maskgct2024}             & 2.27 & 77.4 & 2.62 & 71.4 & 10.27 & 74.8 & 5.05 & 74.5 \\
F5-TTS~\cite{f5tts2025}              & 1.53 & 76.0 & 2.00 & 67.0 & 8.67 & 71.3 & 4.07 & 71.4 \\
DiTAR~\cite{ditar2025}               & 1.02 & 75.3 & 1.69 & 73.5 & - & - & - & - \\
MegaTTS3~\cite{megatts3_2025}            & 1.52 & 79.0 & 2.79 & 77.1 & - & - & - & - \\
MiniMax-Speech~\cite{minimax2025}      & 0.83 & 78.3 & 1.65 & 69.2 & - & - & - & - \\
CosyVoice3-1.5B~\cite{cosyvoice3_2025}     & 1.12 & 78.1 & 2.21 & 72.0 & 5.83 & 75.8 & 3.05 & 75.3 \\
IndexTTS2~\cite{indextts2_2026}           & 1.03 & 76.5 & 2.23 & 70.6 & 7.12 & 75.5 & 3.46 & 74.2 \\
OpenAudio-s1-mini~\cite{openaudios1_2025}   & 1.18 & 68.5 & 1.94 & 55.0 & 23.4 & 64.3 & 8.84 & 62.6 \\
VibeVoice~\cite{vibevoice2025}           & 1.16 & 74.4 & 3.04 & 68.9 & - & - & - & - \\
Qwen3-TTS-12Hz-1.7B-Base~\cite{qwen3tts2026} & 1.22 & 77.0 & \textbf{1.23} & 71.7 & 6.76 & 74.8 & 3.07 & 74.5 \\
MOSS-TTS~\cite{mosstts2026}            & 1.46 & 76.2 & 1.92 & 69.3 & - & - & - & - \\
VoxCPM2~\cite{voxcpm2_2026}             & 0.97 & 79.5 & 1.84 & 75.3 & 8.13 & 75.3 & 3.65 & 76.7 \\
dots.tts~\cite{dotstts2026}            & 0.94 & \textbf{81.0} & 1.30 & 77.1 & 6.60 & \textbf{79.5} & 2.95 & \textbf{79.2} \\
Qwen-Audio-3.0-TTS~\cite{qwenaudio3tts2026}    & 0.84 & 79.2 & 1.54 & 76.2 & 7.00 & 76.8 & 3.13 & 77.4 \\
FireRedTTS3-Base~\cite{fireredtts3_2026}    & 1.01 & 80.9 & 1.64 & \textbf{77.2} & 6.50 & 78.4 & 3.05 & 78.8 \\
\midrule
JoyAI-Voice2.0-Base   & \textbf{0.74} & 77.1 & 1.27 & 73.9 & \textbf{5.53} & 73.4 & \textbf{2.51} & 74.8 \\
\bottomrule
\end{tabular}
}
\end{table}

\subsubsection{Voice Design}

Table~\ref{tab:voice_design_eval} compares JoyAI-Voice2.0-VoiceDesign with leading voice design systems on the InstructTTSEval benchmark. JoyAI-Voice2.0-VoiceDesign achieves the highest APS on both ZH and EN subsets, reaching $90.1$ and $84.4$ respectively, indicating superior adherence to explicit acoustic-parameter instructions. On DSD, JoyAI-Voice2.0-VoiceDesign leads on ZH at $83.3$, and ties with FireRedTTS3-Instruct~\cite{fireredtts3_2026} for the highest EN-DSD at $82.3$. On RP, JoyAI-Voice2.0-VoiceDesign ranks first on ZH at $69.8$, while FireRedTTS3-Instruct leads on EN at $72.0$, substantially ahead of JoyAI-Voice2.0-VoiceDesign at $63.4$, suggesting that instruction adherence for role-play scenarios in English remains an area for improvement. Results for baselines other than MiniMax-VoiceDesign~\cite{minimaxvoicedesign2026}, Qwen3-TTS-12Hz-1.7B-VoiceDesign~\cite{qwen3tts2026}, and VoxCPM2~\cite{voxcpm2_2026} are cited from the FireRedTTS3 technical report~\cite{fireredtts3_2026}, which re-evaluates all systems under Gemini-2.5-Pro following the same judge-substitution protocol as ours, ensuring consistency across the comparison.

\begin{table}[ht]
\centering
\caption{Instruction-following accuracy (\%, $\uparrow$) on InstructTTSEval. APS, DSD, and RP denote the acoustic-parameter specification, descriptive-style directive, and role-play subsets, respectively. Best results are marked in bold.}
\label{tab:voice_design_eval}
\resizebox{\textwidth}{!}{
\begin{tabular}{lcccccc}
\toprule
 & \multicolumn{3}{c}{\textbf{ZH}} & \multicolumn{3}{c}{\textbf{EN}} \\
\cmidrule(lr){2-4} \cmidrule(lr){5-7}
\textbf{Model} & APS ($\uparrow$) & DSD ($\uparrow$) & RP ($\uparrow$) & APS ($\uparrow$) & DSD ($\uparrow$) & RP ($\uparrow$) \\
\midrule
PromptTTS~\cite{prompttts2022}                     & -    & -    & -    & 64.3 & 47.2 & 31.4 \\
PromptStyle~\cite{promptstyle2023}                 & -    & -    & -    & 57.4 & 46.4 & 30.9 \\
VoxInstruct~\cite{voxinstruct2024}                 & 47.5 & 52.3 & 42.6 & 54.9 & 57.0 & 39.3 \\
Parler-tts-mini~\cite{parlertts2024}               & -    & -    & -    & 63.4 & 48.7 & 28.6 \\
Parler-tts-large~\cite{parlertts2024}              & -    & -    & -    & 60.0 & 45.9 & 31.2 \\
MiniMax-VoiceDesign~\cite{minimaxvoicedesign2026}        & 80.1 & 74.6 & 68.7 & 75.5 & 77.4 & 61.8 \\
MiMo-Audio-7B-Instruct~\cite{mimoaudio2025}        & 75.7 & 74.3 & 61.5 & 80.6 & 77.6 & 59.5 \\
VoiceSculptor-VD~\cite{voicesculptor2026}          & 74.6 & 63.5 & 62.0 & -    & -    & -    \\
Qwen3-TTS-12Hz-1.7B-VoiceDesign~\cite{qwen3tts2026} & 85.9 & 81.1 & 62.9 & 77.2 & 80.9 & 63.9 \\
MOSS-VoiceGenerator~\cite{mossvoicegen2026}        & 71.6 & 72.5 & 61.3 & 58.8 & 71.8 & 61.6 \\
VoxCPM2~\cite{voxcpm2_2026}                         & 87.2 & 71.6 & 58.8 & 77.3 & 75.8 & 62.0 \\
Ming-Omni-TTS-16B-A3B~\cite{mingomnitts2026}       & 84.6 & 70.7 & 56.0 & -    & -    & -    \\
FireRedTTS3-Instruct~\cite{fireredtts3_2026}       & 85.8 & 82.0 & 69.7 & 80.7 & \textbf{82.3} & \textbf{72.0} \\
\midrule
JoyAI-Voice2.0-VoiceDesign        & \textbf{90.1} & \textbf{83.3} & \textbf{69.8} & \textbf{84.4} & \textbf{82.3} & 63.4 \\
\bottomrule
\end{tabular}
}
\end{table}

We further evaluate multi-dimensional instruction following on MDVD-Eval. Table~\ref{tab:mdvd_eval} reports per-dimension atomic match rates and their overall average. JoyAI-Voice2.0-VoiceDesign achieves the highest overall score at $0.893$, ahead of VoxCPM2~\cite{voxcpm2_2026} at $0.889$ and Qwen3-TTS-12Hz-1.7B-VoiceDesign~\cite{qwen3tts2026} at $0.849$. At the dimension level, JoyAI-Voice2.0-VoiceDesign leads on 5 of 10 dimensions, with particularly strong advantages on age and gender. Gender achieves near-perfect match rates across most systems, while timbre remains the most challenging dimension overall, indicating a common bottleneck for current voice design systems. MiniMax-VoiceDesign~\cite{minimaxvoicedesign2026} and MiMo-V2.5-TTS-VoiceDesign~\cite{mimov25tts2026} are commercial systems and are evaluated through their official APIs.

\begin{table}[ht]
\centering
\caption{Multi-dimensional instruction following on MDVD-Eval. For each dimension (e.g., Gender, Age, Emotion, Style), we report the atomic match rate ($\uparrow$) between the generated speech attribute and the instructed attribute, along with the overall average across all dimensions. Best results are marked in bold.}
\label{tab:mdvd_eval}
\resizebox{\textwidth}{!}{
\begin{tabular}{lccccccccccc}
\toprule
\textbf{Model} & \textbf{Accent} & \textbf{Age} & \textbf{Gender} & \textbf{Person.} & \textbf{Emotion} & \textbf{Speed} & \textbf{Identity} & \textbf{Timbre} & \textbf{Volume} & \textbf{Style} & \textbf{Overall} \\
\midrule
MiniMax-VoiceDesign~\cite{minimaxvoicedesign2026}        & 0.522 & 0.893 & 0.952 & 0.944 & 0.906 & 0.866 & \textbf{0.897} & 0.700 & 0.776 & \textbf{0.958} & 0.841 \\
MiMo-V2.5-TTS-VoiceDesign~\cite{mimov25tts2026} & 0.560 & 0.874 & \textbf{0.997} & 0.948 & 0.854 & 0.849 & 0.880 & 0.668 & 0.705 & 0.916 & 0.825 \\
Qwen3-TTS-12Hz-1.7B-VoiceDesign~\cite{qwen3tts2026}                  & 0.609 & 0.906 & 0.971 & 0.956 & 0.940 & 0.920 & 0.806 & 0.685 & 0.805 & 0.891 & 0.849 \\
VoxCPM2~\cite{voxcpm2_2026}                    & 0.723 & 0.918 & 0.981 & 0.952 & \textbf{0.955} & \textbf{0.952} & 0.855 & \textbf{0.743} & 0.879 & 0.933 & 0.889 \\
\midrule
JoyAI-Voice2.0-VoiceDesign   & \textbf{0.741} & \textbf{0.953} & \textbf{0.997} & \textbf{0.972} & 0.944 & 0.915 & 0.880 & 0.739 & \textbf{0.885} & 0.908 & \textbf{0.893} \\
\bottomrule
\end{tabular}
}
\end{table}

\section{Conclusion}
\label{sec:conclusion}

This report presented JoyAI-Voice~2.0, an end-to-end anthropomorphic speech generation model built on a fully continuous dual-encoder architecture. Departing from the discrete-token paradigm, JoyAI-Voice~2.0 removes the quantization information bottleneck entirely. The AudioVAE compresses speech into continuous latents, and the semantic-acoustic dual encoder decomposes each patch into a purified semantic summary and an information-dense acoustic aggregation. Built upon the semantic-acoustic joint representation, a 1.5B causal language model performs long-range semantic-prosodic planning over the semantic stream, while a local flow-matching diffusion Transformer renders full 48\,kHz waveforms patch by patch. This explicit representation-level separation shields the language model from acoustic noise propagation and closes the error-feedback loop that plagues open-loop continuous rollouts, yielding rollout stability approaching that of discrete systems without sacrificing continuous fidelity.

Comprehensive experiments validate the effectiveness of the design. On the Seed-TTS benchmark, JoyAI-Voice2.0-Base achieves the lowest average recognition error of 2.51\%, a 14.9\% relative reduction over the strongest baseline, with leading robustness on challenging utterances. The voice design variant attains state-of-the-art attribute fidelity on InstructTTSEval in both Chinese and English, and ranks first overall on MDVD-Eval, our proposed multi-dimensional benchmark, leading on 5 of 10 perceptual dimensions. Beyond these benchmarks, JoyAI-Voice~2.0 supports instruction-based voice design, zero-shot voice cloning robust to noisy references, and controllable paralinguistic tags such as \texttt{[Laughter]} and \texttt{[Sigh]}. Together with streaming inference at low first-packet latency, these capabilities make it well suited for anthropomorphic, empathetic, and real-time conversational deployment.

\section*{Authors (alphabetical order of family name)}

\begin{description}
    \item[Core Contributors:] Yafeng Chen, Boya Dong, Yankun Huang, Hao Li, Jingdong Li, Xiangyu Liang, Hao Ni, Wenchao Wang, Yuxuan Wang, Zhangyu Xiao
    \item[Contributors:] Wei Deng, Nan Duan, Yu Gu, Wenhao Guan (Intern), Weisheng Han, Yabin Li, Yuan Liu, Jiaxin Ye (Intern), Fan Yu, Lin Zhu
\end{description}



\bibliographystyle{unsrt}
\bibliography{references}

\end{document}